\documentclass[letterpaper,11pt]{article}

\usepackage[tbtags]{amsmath}
\usepackage{amssymb,amsfonts,amsthm}
\usepackage{graphicx}
\usepackage{setspace}
\usepackage{xspace}
\usepackage{natbib}
\usepackage{dsfont} 
\usepackage{morefloats}
\usepackage{verbatim}

\newcommand{\unit}[1]{\ensuremath{\, \mathrm{#1}}}
\newcommand{\rd}{\textrm{d}}
\newcommand{\R}{\mathbb{R}}
\newcommand{\ldef}{\mathrel{:=}}
\newcommand{\iid}{\buildrel \rm iid \over \sim}
\newcommand{\E}{\mathop{\textrm{E}}}
\newcommand{\Var}{\mathop{\textrm{Var}}}
\renewcommand{\Pr}{\mathop{\textrm{Pr}}}

\newcommand{\indic}{\mathds{1}}

\renewcommand{\thetable}{\Roman{table}}
\renewcommand{\refname}{REFERENCES} 

\title{Rapidly bounding the exceedance probabilities of high aggregate losses}

\author{Isabella Gollini\footnote{Corresponding author. email: isabella.gollini@bristol.ac.uk}\\Department of Engineering\\University of Bristol
 \and Jonathan Rougier\\Department of Mathematics\\University of Bristol}

\date{\today}

\begin{document}

\maketitle


\begin{abstract}\noindent%
  We consider the task of assessing the righthand tail of an insurer's
  loss distribution for some specified period, such as a year.  We
  present and analyse six different approaches: four upper bounds,
  and two approximations.  We examine these approaches under a variety
  of conditions, using a large event loss table for US hurricanes.
  For its combination of tightness and computational speed, we favour
  the Moment bound.  We also consider the appropriate size of Monte
  Carlo simulations, and the imposition of a cap on single event
  losses.  We strongly favour the Gamma distribution as a flexible
  model for single event losses, for its tractable form in all of the
  methods we analyse, its generalisability, and because of the ease
  with which a cap on losses can be incorporated.

\smallskip \noindent%
\end{abstract}

\newpage

\section{\MakeUppercase{Introduction}}
\label{sec:introduction}

One of the objectives in catastrophe modelling is to assess the
probability distribution of losses for a specified period, such as a
year.  From the point of view of an insurance company, the whole of
the loss distribution is interesting, and valuable in determining
insurance premiums.  But the shape of the righthand tail is critical,
because it impinges on the solvency of the company.  A simple measure
of the risk of insolvency is the probability that the annual loss will
exceed the company's current operating capital.  Imposing an upper
limit on this probability is one of the objectives of the EU
Solvency~II directive.

If a probabilistic model is supplied for the loss process, then this
tail probability can be computed, either directly, or by simulation.
\citet{shev10} provides a survey of the various approaches.  This can
be a lengthy calculation for complex losses.  Given the inevitably
subjective nature of quantifying loss distributions, computational
resources might be better used in a sensitivity analysis.  This requires
either a quick approximation to the tail probability or an upper bound
on the probability, ideally a tight one.  In this paper we present
and analyse several different bounds, all of which can be computed quickly from a
very general event loss table.  
By making no assumptions about the shape of the righthand tail beyond the existence of the second moment, our approach extends to fat-tailed distributions.
We provide a numerical illustration,
and discuss the conditions under which the bound is tight.

 \section{\MakeUppercase{Interpreting the Event Loss Table}}
\label{sec:ELT}

We use a rather general form for the Event Loss Table (ELT), given in
Table~\ref{fig:ELT}.  In this form, the losses from an identified
event $i$ are themselves uncertain, and described by a probability
density function $f_i$.  That is to say, if $X_i$ is the loss from a
single occurrence of event $i$, then
\begin{displaymath}
  \Pr(X_i \in A) = \int_A f_i(x) \, \rd x
\end{displaymath}
for any well-behaved $A \subset \R$.  The special case where the loss for an occurrence of
event $i$ is treated as a constant $x_i$ is represented with the Dirac
delta function $f_i(x) = \delta(x - x_i)$.

The choice of $f_i$ for each event represents represents uncertainty about the loss that follows from the event, often termed `secondary uncertainty' in catastrophe modelling.  We will discuss an efficient and flexible approach to representing more-or-less arbitrary specifications of $f_i$ in Section~\ref{sec:special}.

\begin{table}\renewcommand{\arraystretch}{1.25}
  \centering
  \caption{Generic Event Loss Table (ELT).  Row $i$ represents an
    event with arrival rate $\lambda_i$, and loss distribution
    $f_i$.}
  \label{fig:ELT}
  \bigskip
  \begin{tabular}{*{3}{c}}
    \hline
    Event ID & Arrival rate, $\unit{yr}^{-1}$ & Loss distribution \\
    \hline
    1 & $\lambda_1$ & $f_1$ \\
    2 & $\lambda_2$ & $f_2$ \\
    $\vdots$ & $\vdots$ & $\vdots$ \\
    $m$ & $\lambda_m$ & $f_m$ \\
    \hline\hline
  \end{tabular}
\end{table}

There are two equivalent representations of the ELT, for stochastic
simulation of the loss process through time \citep[see,
e.g.,][sec.~1.5]{ross96}.  The first is that the $m$ events with
different IDs follow concurrent but independent homogeneous Poisson
processes.  The second is that the collective of events follows a
single homogeneous Poisson process with arrival rate
\begin{displaymath}
  \lambda \ldef \sum_{i=1}^m \lambda_i 
\end{displaymath}
and then, when an event occurs, its ID is selected independently at
random with probability $\lambda_i / \lambda$.

The second approach is more tractable for our purposes.  Therefore we
define $Y$ as the loss incurred by a randomly selected event, with
probability density function
\begin{displaymath}
  f_Y = \sum_{i=1}^m \frac{ \lambda_i}{ \lambda} f_i \, .
\end{displaymath}
The total loss incurred over an interval of length $t$ is then
modelled as the random sum of independent losses, or
\begin{displaymath}
  S_t \ldef \sum_{j=1}^{N_t} Y_j \qquad \text{where} \quad
  \begin{cases}
    N_t \sim \text{Poisson}(\lambda t) \text{, and} \\
    Y_1, Y_2, \dots \iid f_Y .
  \end{cases}
\end{displaymath}
The total loss $S_t$ would generally be termed a compound Poisson
process with rate $\lambda$ and component distribution $f_Y$.  An
unusual feature of loss modelling is that the component distribution
$f_Y$ is itself a mixture, sometimes with thousands of components.

%
%
  
 \section{\MakeUppercase{A selection of upper bounds}}
\label{sec:bounds}

Our interest is in a bound for the probability $\Pr(S_t \geq s)$ for
some specified ELT and time period $t$; we assume, as is natural, that
$\Pr(S_t \leq 0) = 0$.  We pose the question: is $\Pr(S_t \geq s)$
small enough to be tolerable for specified $s$ and $t$?  We are aware
of four useful upper bounds on $\Pr(S_t \geq s)$, explored here in
terms of increasing complexity.  The following material is covered in
standard textbook such as \citet{grimmett01}, and in more specialised
books such as \citet{ross96} and \citet{whittle00}.  To avoid clutter,
we will drop the `$t$' subscript on $S_t$ and $N_t$.

\paragraph{The Markov inequality.}

The Markov inequality states that if ${\Pr(S \leq 0) = 0}$ then
\begin{equation}
  \label{eq:Mar}
  \Pr( S \geq s ) \leq \frac{ \mu }{ s } \tag{Mar}
\end{equation}
where $\mu \ldef \E(S)$.  As $S$ is a compound process,
\begin{equation}
  \label{eq:Smom1}
  \mu = \E(N) \E(Y) = \lambda t \E(Y) ,
\end{equation}
the second equality following because $N$ is Poisson.  The second
expectation is simply
\begin{displaymath}
  \E(Y) = \sum_{i=1}^m \frac{\lambda_i}{\lambda} \E(X_i) .
\end{displaymath}
We do not expect this inequality to be very tight, because it imposes
no conditions on the integrability of $S^2$, but it is so fast to
compute that it is always worth a try for a large $s$.

\paragraph{The Cantelli inequality.}

If $S$ is square-integrable, i.e.\ $\sigma^2 \ldef \Var(S)$ is finite,
then
\begin{equation}
  \label{eq:Cant}
  \Pr( S \geq s) \leq \frac{ \sigma^2 }{ \sigma^2 + (s - \mu)^2 } \quad \text{for $s \geq \mu$.}
  \tag{Cant}
\end{equation}
This is the Cantelli inequality, and it is derived from the Markov
inequality.  As $S$ is a compound process,
\begin{equation}
  \label{eq:Smom2}
  \sigma^2 = \E(N) \Var(Y) + \E(Y)^2 \Var(N) = \lambda t \E(Y^2) ,
\end{equation}
the second equality following because $N$ is Poisson.  The second
expectation is simply
\begin{displaymath}
  \E(Y^2) = \sum_{i=1}^m \frac{\lambda_i}{\lambda} \E(X_i^2) .
\end{displaymath}

We expect the Cantelli bound will perform much better than the Markov
bound both because it exploits the fact that $S$ is square integrable,
and because its derivation involves an optimisation step.  It is
almost as cheap to compute, and so it is really a free upgrade.

\paragraph{The moment inequality.}

This inequality and the Chernoff inequality below use the generalised
Markov inequality: if $g$ is increasing, then $S \geq s \iff g(S) \geq
g(s)$, and so
\begin{displaymath}
  \Pr( S \geq s ) \leq \frac{ \E\{g(S)\} }{ g(s) }
\end{displaymath}
for any $g$ that is increasing and non-negative.

An application of the generalised Markov inequality gives
\begin{displaymath}
  \Pr(S \geq s) \leq \inf_{k > 0} \frac{ \E(S^k) }{ s^k } , 
\end{displaymath}
because $g(s) = s^k$ is non-negative and increasing for all $k > 0$.
Fractional moments can be tricky to compute, but integer moments are
possible for compound Poisson processes.  Hence we consider
\begin{equation}
  \label{eq:Mom}
  \Pr(S \geq s) \leq \min_{k = 1, 2 \dots} \frac{ \E(S^k) }{ s^k } . \tag{Mom}
\end{equation}
This cannot do worse that the Markov bound, which is the special case
of $k = 1$.

The integer moments of a compound Poisson process can be computed
recursively, as shown in \citet[sec.~2.5.1]{ross96}:
\begin{equation}
  \label{eq:cpp}
  \E(S^k) = \lambda t \sum_{j=0}^{k-1} {k - 1 \choose j} \E(S^j) \E(Y^{k-j}) .
\end{equation}
The only new term here is
\begin{displaymath}
  \E(Y^{k-j}) = \sum_{i=1}^m \frac{ \lambda_i }{ \lambda } \E(X_i^{k-j}) .
\end{displaymath}
At this point it would be helpful to know the Moment Generating
Function (MGF, see below) of each $X_i$.

Although not as cheap as the Cantelli bound, this does not appear to
be an expensive calculation, if the $f_i$'s have standard forms with
simple known MGFs.  It is legitimate to stop at any value of $k$, and
it might be wise to limit $k$ in order to avoid numerical issues with
sums of very large values.

\paragraph{The Chernoff inequality.}

Let $M_S$ be the MGF of $S$, that is
\begin{displaymath}
  M_S(v) \ldef \E\big( e^{v S} \big) \qquad v \geq 0 . 
\end{displaymath}
Chernoff's inequality states
\begin{equation}
  \label{eq:Ch}
  \Pr(S \geq s) \leq \inf_{k > 0} \, \frac{ M_S(k) }{ e^{k s} }. \tag{Ch}
\end{equation}
It follows from the generalised Markov inequality with $g(s) = e^{k s}$,
which is non-negative and increasing for all $k > 0$.

If $M_Y$ is the MGF of $Y$, then
\begin{displaymath}
  M_S(v) = M_N(\log M_Y(v)) \qquad v \geq 0 .
\end{displaymath}
In our model $N$ is Poisson, and hence
\begin{displaymath}
  M_N(v) = \exp\big\{ \lambda t (e^v - 1) \big\} \qquad v \geq 0
\end{displaymath}
\citep[see, e.g.][sec.~1.4]{ross96}.  Thus the MGF of $S$ simplifies
to
\begin{displaymath}
  M_S(v) = \exp\big\{ \lambda t \big( M_Y(v) - 1 \big) \big\}.
\end{displaymath}
The MGF of $Y$ can be expressed in terms of the MGFs of the $X_i$'s:
\begin{displaymath}
  M_Y(v) = \sum_{i=1}^m \frac{ \lambda_i }{ \lambda } M_{X_i}(v) .
\end{displaymath}
Now it is crucial that the $f_i$ have standard forms with simple known
MGFs.

In an unlimited optimisation, the Chernoff bound will never outperform
the Moment bound \citep{philips95}.  In practice, however, constraints
on the optimisation of the Moment bound may result in the best
available Chernoff bound being lower than the best available Moment
bound.  But there is another reason to include the Chernoff bound,
from large deviation theory; see, e.g., \citet[sec.~15.6 and
ch.~18]{whittle00}.  Let $t$ be an integer number of years, and define
$S_1$ as the loss from one year, so that $M_{S_t}(k) = \{ M_{S_1}(k)
\}^t$.  Then large deviation theory states that
\begin{displaymath}
  \Pr(S_t \geq s) = \inf_{k > 0} \exp \big\{ {-k s} + t \log M_{S_1}(k) + o(t) \big\}
\end{displaymath}
and so as $t$ becomes large the Chernoff upper bound becomes exact.
Very informally, then, the convergence of the Chernoff bound and the
Moment bound suggest, according to a squeezing argument, that both
bounds are converging from above on the actual probability.

 \section{\MakeUppercase{Two `exact' approaches}}
\label{sec:exact}

There are several approaches to computing $\Pr(S_t \geq s)$ to arbitrary
accuracy, although in practice this accuracy is limited by computing
power \citep[see][for a review]{shev10}.  We mention two here.

\paragraph{Monte Carlo simulation.}

One realisation of $S_t$ for a fixed time-interval can be generated by
discrete event simulation, also known as the Gillepsie algorithm
\citep[see, e.g.,][sec.~6.4]{wilkinson12}.  Many such simulations can
be used to approximate the distribution function of $S_t$, and can be
used to estimate probabilities, including tail probabilities.

Being finite-sample estimates, these probabilities should have a
measure of uncertainty attached.  This is obviously an issue for
regulation, where the requirement is often to demonstrate that
\begin{displaymath}
  \Pr(S_1 \geq s_0) \leq \kappa_0
\end{displaymath}
for some $s_0$ which reflects the insurer's available capital, and
some $\kappa_0$ specified by the regulator.  For Solvency~II,
$\kappa_0 = 0.005$ for one-year total losses.  A Monte Carlo point
estimate of $p_0 \ldef \Pr(S_1 \geq s_0)$ which was less than
$\kappa_0$ would be much more reassuring if the whole of the 95\%
confidence interval for $p_0$ were less than $\kappa_0$, than if the
95\% confidence interval contained $\kappa_0$.

A similar problem is faced in ecotoxicology, where one recommendation
would be equivalent in this context to requiring that the upper bound
of a 95\% confidence interval for $p_0$ is no greater than $\kappa_0$;
see \citet{hickey13}.  If we adopt this approach, though, it is
incorrect simply to monitor the upper bound and stop sampling when it
drops below $\kappa_0$, because the confidence interval in this case
ought to account for the stochastic stopping rule, rather than being
based on a fixed sample size.  But it is possible to do a design
calculation to suggest an appropriate value for $n$, the sample size,
that will ensure that the upper bound will be larger than $\kappa_0$
with specified probability, \textit{a priori}, as we now discuss.

Let $u_{1-\alpha}(x; n)$ be the upper limit of a level $(1 - \alpha)$
confidence interval for $p_0$, where $x$ is the number of sample
members that are at least $s_0$, and $n$ is the sample size.  Suppose
that the \textit{a priori} probability of this upper limit being no
larger than $\kappa_0$ is to be at least $\beta_0$, where $\beta_0$
would be specified.  In that case, valid $n$'s satisfy
\begin{displaymath}
  \Pr \big\{ u_{1-\alpha}(X; n) \leq \kappa_0 \big\} \geq \beta_0 
\end{displaymath}
where $X \sim \text{Binom}(n, p_0)$.  

\begin{figure}
  \centering
  \includegraphics{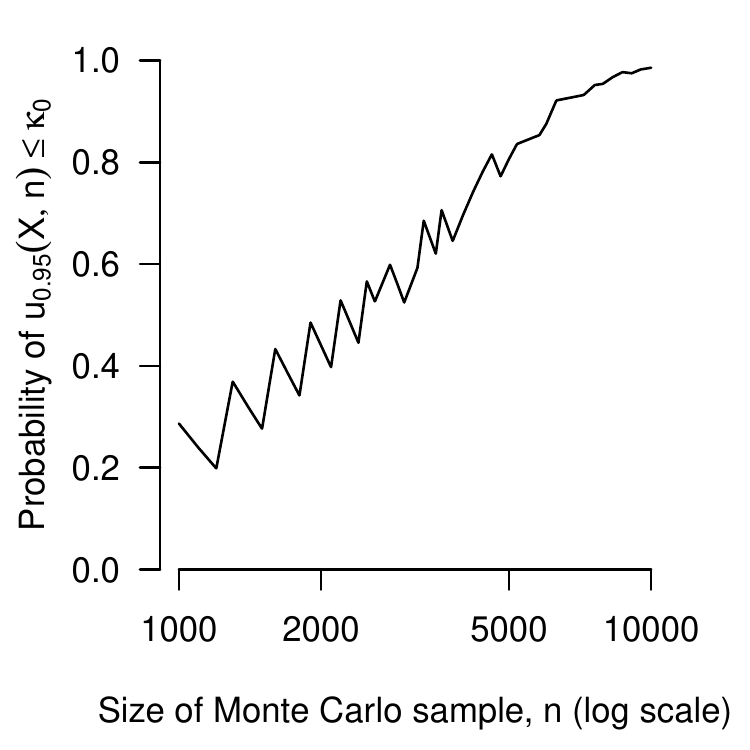}
  \caption{The effect of sample size on Monte Carlo accuracy.  The
    graph shows the probability that the upper bound of the $95\%$
    Jeffreys confidence interval for $p_0$ lies below $\kappa_0 =
    0.005$ when $p_0 = \kappa_0 / 2$.}
  \label{fig:findn}
\end{figure}

There are several ways of constructing an approximate $(1-\alpha)$
confidence interval for $p_0$, reviewed in
\citet{brown01}.\footnote{It is not possible to construct an exact
  confidence interval without using an auxiliary randomisation.}  We
suggest what they term the (unmodified) Jeffreys confidence interval,
which is simply the equi-tailed $(1 - \alpha)$ credible interval for
$p_0$ with the Jeffreys prior, with a minor modification.  Using this
confidence interval, Figure~\ref{fig:findn} shows the probability for
various choices of $n$ with $\kappa_0 = 0.005$ and $p_0 = \kappa_0 /
2$.  In this case, $n = 10^5$ seems to be a good choice, and this
number is widely used in practice.

\paragraph{Panjer recursion.}

The second approach is Panjer recursion; see
\citet[Cor.~2.5.4]{ross96} or \citet[sec.~5]{shev10}.  This provides a
recursive calculation for $\Pr(S_t = s)$ whenever each $X_i$ is
integer-valued, so that $S$ itself is integer-valued.  This
calculation would often grind to a halt if applied literally, but can
be used to provide an approximation if the ELT is compressed, as
discussed in section~\ref{sec:merging}.

Perhaps the main difficulty with Panjer recursion, once it has been
efficiently encoded, is that it does not provide any assessment of the
error which follows from the compression of the ELT.  In this
situation, a precise and computationally cheap upper bound may be of
more practical use than an approximation.  Section~\ref{sec:merging}
also discusses indirect ways to assess the compression error, using
the upper bounds.

Monte Carlo simulation is an attractive alternative to Panjer
recursion, because it comes with a simple assessment of accuracy,
is easily parallelisable, and the sample drawn can be used to calculated other quantities of interest for insurers like the net aggregate loss and reinsurance recovery costs.

 \section{\MakeUppercase{Tractable special cases}}
\label{sec:special}

In this section we consider three tractable special cases.

First, suppose that
\begin{displaymath}
   f_i(x) = \delta(x - x_i) \qquad i = 1, \dots, m,
\end{displaymath}
i.e.\ the loss from event $i$ is fixed at $x_i$.  Then
\begin{displaymath}
  \E(X_i^k) = x_i^k \quad \text{and} \quad M_{X_i}(v) = e^{v x_i} .
\end{displaymath}
All of the bounds are trivial to compute.

Second, suppose that each $f_i$ is a Gamma distribution with
parameters $(\alpha_i, \beta_i)$:
\begin{displaymath}
  f_i(x) = \text{Gam}(x; \alpha_i , \beta_i)  = \frac{ \beta_i^{\alpha_i} } {\Gamma( \alpha_i) } x^{\alpha_i - 1} e^{-\beta_i x} \indic_{x > 0}
  \qquad i = 1, \dots, m
\end{displaymath}
for $\alpha_i, \beta_i > 0$, where $\indic$ is the indicator function and
$\Gamma$ is the Gamma function,
\begin{displaymath}
  \Gamma(s) \ldef \int_0^\infty x^{s-1} e^{-x} \, \rd x .
\end{displaymath}
Then
\begin{equation}
  \label{eq:MGFgamma}
  M_i(v) = \left( \frac{ \beta_i }{ \beta_i - v } \right)^{\alpha_i} \qquad 0
  \leq v < \beta_i .
\end{equation}
The moments are
\begin{equation}
  \label{eq:gammamom}
  \E(X_i^k) = \frac{ \Gamma(\alpha_i + k) }{ \beta_i^k \, \Gamma(\alpha_i) }
\end{equation}
and hence
\begin{displaymath}
  \E(X_i) = \frac{ \alpha_i }{ \beta_i } , \quad \E(X_i^2) = \frac{ (\alpha_i + 1) \alpha_i }{ \beta_i^2 } .
\end{displaymath}

Third, suppose that each $f_i$ is a finite mixture of Gamma
distributions:
\begin{displaymath}
  f_i(x) = \sum_{k=1}^{p_i} \pi_{ik} \, \text{Gam}(x; \alpha_{ik} , \beta_{ik})
  \qquad i = 1, \dots, m
\end{displaymath}
where $\sum_{k=1}^{p_i} \pi_{ik} = 1$ for each $i$.  Then
\begin{align*}
  f_Y(y) 
  & = \sum_{i=1}^m \frac{ \lambda_i}{ \lambda} \sum_{k=1}^{p_i} \pi_{ik} \,
  \text{Gam}(y; \alpha_{ik} , \beta_{ik}) \\
  & = \sum_{i=1}^m \sum_{k=1}^{p_i} \frac{ \lambda_i \pi_{ik}}{ \lambda}
  \text{Gam}(y; \alpha_{ik} , \beta_{ik}) .
\end{align*}
In other words, this is exactly the same as creating an extended ELT
with plain Gamma $f_i$'s (i.e.\ as in the second case), but where each
$\lambda_i$ is shared out among the $p_i$ mixture components according
to the mixture weights $\pi_{i1}, \dots, \pi_{ip_i}$.

This third case is very helpful, because the Gamma calculation is so simple, and yet it is possible to approximate any strictly positive absolutely continuous probability density function that has limit zero as $x \to\infty$, with a mixture of Gamma distributions \citep{Wiper01}. ÊIt is also possible to approximate point distributions by very concentrated Gamma distributions, discussed below in Section~\ref{sec:thickening}. ÊThus the secondary uncertainty for an event might be represented as a set of discrete losses, each with its own probability, but encoded as a set of highly concentrated Gamma distributions, leading to very efficient calculations.


\paragraph{Capped single-event losses.}

For insurers, a rescaled Beta distribution is often preferred to a
Gamma distribution, because it has a finite upper limit representing
the maximum insured loss.  The moment generating function of a Beta
distribution is an untabulated function with an infinite series
representation, and so will be more expensive to compute accurately;
this will affect the Chernoff bound.  There are no difficulties with
the moments.  

However, we would question the suitability of using a Beta
distribution here.  The insurer's loss from an event is capped at the
maximum insured loss.  This implies an atom of probability at the
maximum insured loss: if $f_i$ is the original loss distribution for
event $i$ and $u$ is the maximum insured loss, then
\begin{displaymath}
  f_i(x; u) = f_i(x) \indic_{x < u} + (1-p_i) \delta(x - u)
\end{displaymath}
where $p_i \ldef \int_0^{u} f_i(x) \, \rd x$ and $\delta$ is the
Dirac delta function, as before.  A Beta distribution scaled to $[0,
u]$ would be quite different, having no atom at $u$.

The Gamma distribution for $f_i$ is tractable with a cap on losses.
If $f_i$ is a Gamma distribution then the MGF is
\begin{displaymath}
  M_i(v; u) = \left( \frac{ \beta_i }{ \beta_i - v } \right)^{\alpha_i} \frac{ \gamma(\alpha_i, (\beta_i - v) u) }{ \Gamma(\alpha_i) } + (1 - p_i) e^{v u} ,
\end{displaymath}
where
$\gamma$ is the incomplete Gamma function,
\begin{displaymath}
  \gamma(s, u) \ldef \int_0^u x^{s-1} e^{-x} \, \rd x ,
\end{displaymath}
and 
\begin{displaymath}
  p_i \ldef \frac{ \gamma(\alpha_i, \beta_i u) }{ \Gamma(\alpha_i) } .
\end{displaymath}
The moments of $f_i(\cdot; u)$ are
\begin{displaymath}
  \E(X_i^k; u) = \frac{ \gamma(k + \alpha_i, \beta_i u) }{ \beta_i^k \, \Gamma(\alpha_i)} + (1 - p_i) u^k
\end{displaymath}
Introducing a non-zero lower bound is straightforward.

 \section{\MakeUppercase{Numerical Illustration}}
\label{sec:numerical}

We have implemented the methods of this paper in a package for the R
open source statistical computing environment \citep{Rproject}, named
\texttt{tailloss}.  In addition, this package includes a large ELT for
US hurricanes (32,060 rows).

\subsection{The effect of merging}
\label{sec:merging}

We provide a utility function, \texttt{compressELT}, which reduces the
number of rows of an ELT by rounding and merging.  This speeds up all
of the calculations, and is crucial for the successful completion of
the Panjer approximation.

The rounding operation rounds each of the losses to a specified number
$d$ of decimal places, with $d = 0$ being to an integer, and $d < 0$
being a value with $d$ zeros before the decimal point.  Then the
rounded value is multiplied by $10^d$ to convert it to an integer.
Finally, the merge operation combines all the rows of the ELT with the
same transformed loss, and adds their rates.

\begin{table}\renewcommand{\arraystretch}{1.25}
  \centering
  \caption{ELT US Hurricane dataset.  Row $i$ represents an
    event with arrival rate $\lambda_i$, and expected loss
    $x_i$.}
 \label{tab:ELTcompd0}
 \bigskip
  \begin{tabular}{*{3}{r}}
  \hline
 Event ID & Arrival rate, $\unit{yr}^{-1}$ & Expected Loss, \$ \\
    \hline
    1 &    0.09265 &          1 \\ 
2 &    0.03143 &          2 \\ 
3 &    0.02159 &          3 \\ 
4 &    0.01231 &          4 \\ 
5 &    0.01472 &          5 \\ 
$\vdots$ & $\vdots$ & $\vdots$ \\ 
32056 &    0.00001 &   17593790 \\ 
32057 &    0.00001 &   18218506 \\ 
32058 &    0.00001 &   18297003 \\ 
32059 &    0.00001 &   19970669 \\ 
32060 &    0.00001 &   24391615 \\ 

   \hline\hline
\end{tabular}
\end{table}

\begin{table}\renewcommand{\arraystretch}{1.25}
  \centering
\caption{ELT US Hurricane dataset, after rounding and merging  to  \$10k ($d = {-4}$). \textit{Cf.}~Table~\ref{tab:ELTcompd0}.}
 \label{tab:ELTcompd-4}
 \bigskip
  \begin{tabular}{*{3}{r}}
  \hline
 Event ID & Arrival rate, $\unit{yr}^{-1}$ & Expected Loss, \$10k \\
    \hline
    1 &    0.35764 &          1 \\ 
2 &    0.16864 &          2 \\ 
3 &    0.16088 &          3 \\ 
4 &    0.12135 &          4 \\ 
5 &    0.12239 &          5 \\ 
$\vdots$ & $\vdots$ & $\vdots$ \\ 
1141 &    0.00001 &       1759 \\ 
1142 &    0.00001 &       1822 \\ 
1143 &    0.00001 &       1830 \\ 
1144 &    0.00001 &       1997 \\ 
1145 &    0.00001 &       2439 \\ 

   \hline\hline
\end{tabular}
\end{table}

Table~\ref{tab:ELTcompd0} shows some of the original ELT, and
Table~\ref{tab:ELTcompd-4} the same table after rounding to the
nearest \$10k (i.e.\ ${d = {-4}}$).  It is an empirical question, how
much rounding can be performed on a given ELT without materially
changing the distribution of $t$-year total losses.  Ideally, this
would be assessed using an exact calculation, like Panjer recursion.
Unfortunately it is precisely because Panjer recursion is so
numerically intensive that rounding and merging of large ELTs is
necessary in the first place.  So instead we assess the effect of
rounding and merging using the Moment bound, which, as already
established, converges to the actual value when the number of events
in the time-interval is large.

\begin{figure}[p]
  \includegraphics[width=\textwidth]{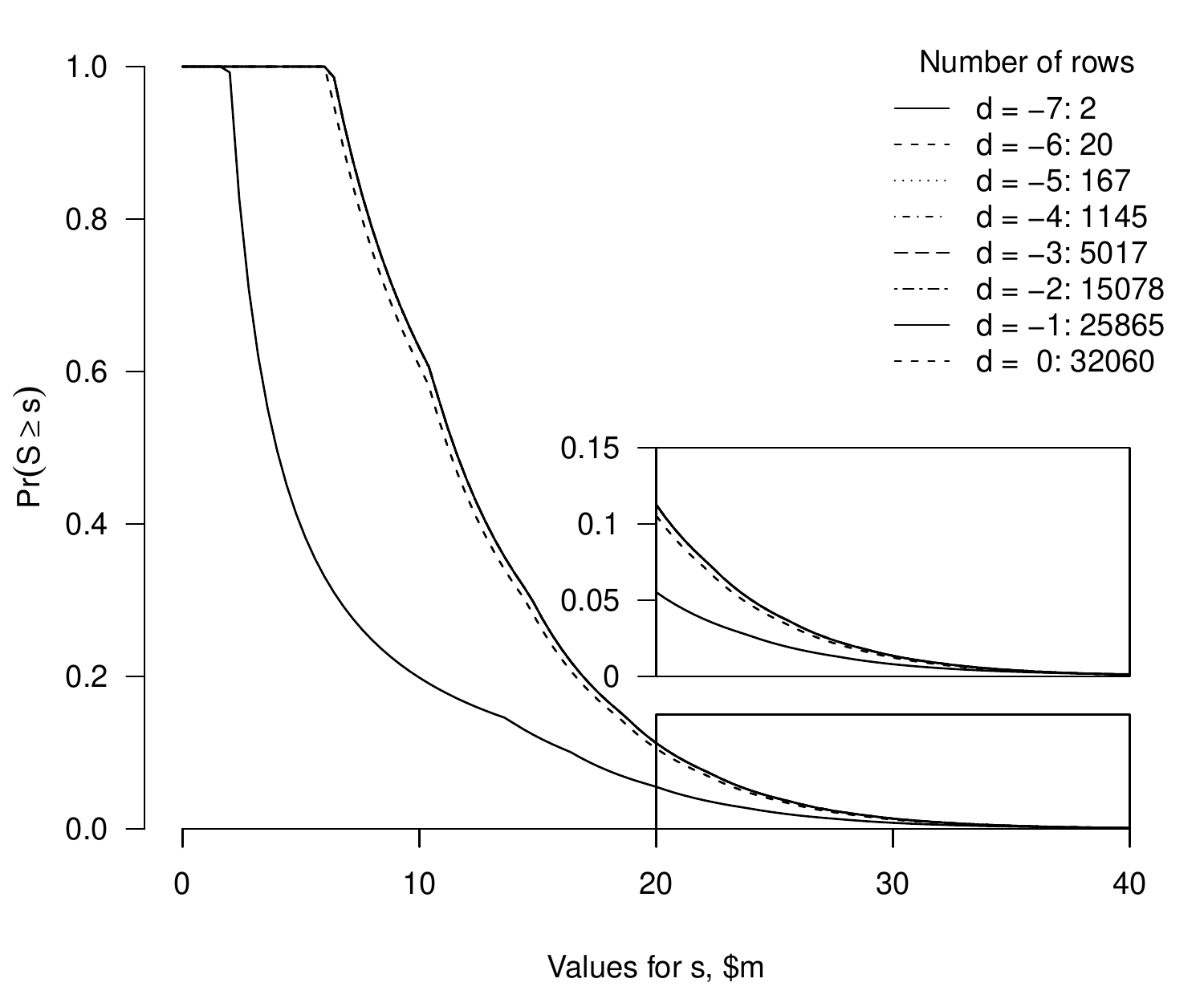}
  \caption{The effect of compression and merging on the US Hurricanes
    ELT.  The curves show the values of the Moment bound on the
    exceedance probability for one-year total losses.  All values of
    $d$ larger than $-7$ (only two rows) give very similar outcomes,
    with values of $-5$ or larger being effectively identical, and
    overlaid on the Figure.}
  \label{fig:effectOfCompression}
\end{figure}

Figure~\ref{fig:effectOfCompression} shows the result of eight
different values for $d$, from ${-7}$ to $0$.  The outcome with ${d =
  {-7}}$ is materially different, which is not surprising because this
ELT only has two rows.  More intriguing is that the outcome with ${d =
  {-6}}$ is almost the same as that with no compression at all,
despite the ELT having only $20$ rows.

\subsection{Computational expense of the different methods}
\label{sec:timings}

Here we consider one-year losses, and treat the losses for each event
as certain; i.e.\ the first case in section~\ref{sec:special}.  The
methods we consider are Panjer, Monte~Carlo, Moment, Chernoff,
Cantelli, and Markov.  The first two provide approximately exact
values for $\Pr(S_1 \geq s)$.  Panjer is an approximation because of
the need to compress the ELT.  For the Monte Carlo method, we used
$10^5$ simulations, as discussed in section~\ref{sec:exact}, and we
report the $95\%$ confidence interval in the tail.  The remaining
methods provide strict upper bounds on $\Pr(S_1 \geq s)$.  Our
optimisation approach for the Moment and Chernoff bounds is given in
the Appendix. 
All the timings are CPU times in seconds on a MacBook Pro processor 2.53 GHz Intel Core 2 Duo.

\begin{figure}[p]
  \includegraphics[width=\textwidth]{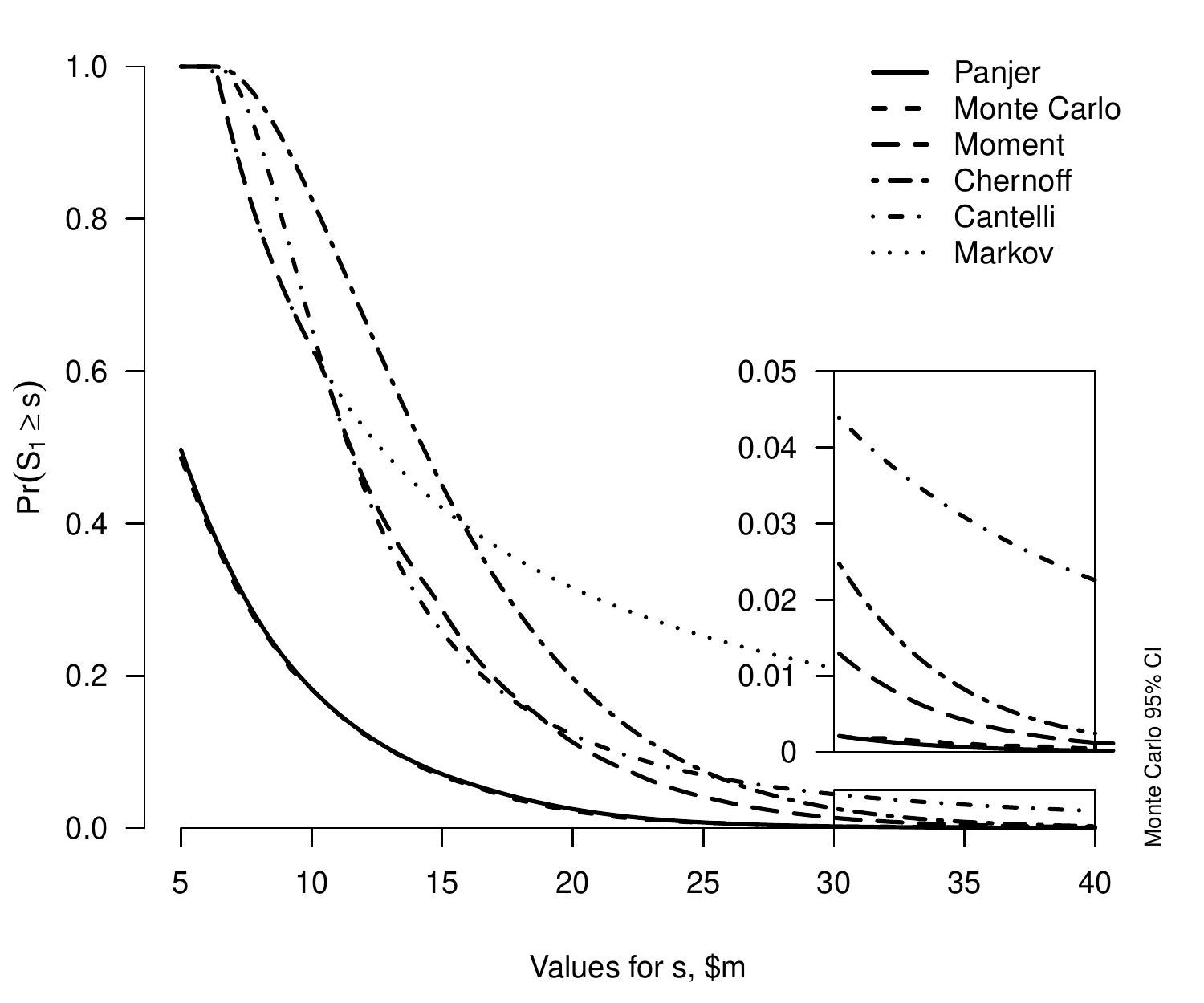}
  \caption{Exceedance probabilities for the methods, with rounding of
    ${d=-4}$ on the US Hurricanes ELT.  The legend shows the Monte
    Carlo 95\% confidence interval for $p_0$ at $s_0 = \$40 \unit{m}$;
    see section~\ref{sec:exact}.  Each curve comprises 101 points,
    equally-spaced between $\$0 \unit{m}$ and $\$40 \unit{m}$.
    Timings are given in Table~\ref{tab:timingsA}.  For later reference,
    this Figure has $t = 1$, $u = \infty$, $\theta = 0$, and $d =
    -4$.}
  \label{fig:EPA}
\end{figure}

Figure~\ref{fig:EPA} shows the exceedance probabilities for the
methods, computed on $101$ equally-spaced ordinates between $\$0
\unit{m}$ and $\$40 \unit{m}$, with compression $d = -4$.  The Markov
bound is the least effective, and the Cantelli bound is surprisingly
good.  As expected, the Chernoff and Moment bounds converge, and also,
in this case, converge on the Panjer and Monte Carlo estimates.

The timings for the methods are given in Table~\ref{tab:timingsA}.
These values require very little elaboration.  The Moment, Cantelli,
and Markov bounds are effectively instantaneous to compute, with
timings of a few thousandths of a second.  The Chernoff bound is more
expensive but still takes only a fraction of a second.  The Monte
Carlo and Panjer approximations are hundreds, thousands, or even
millions of times more expensive.  The Panjer bound is impractical to
compute at compression below ${d = -2}$ (and from now on we will just
consider $d \leq -3$).

\begin{table}
  \centering\renewcommand{\arraystretch}{1.25}
  \caption{Timings for the  methods shown in Figure~\ref{fig:EPA}, in seconds
    on a standard desktop computer, for different degrees of rounding (see section~\ref{sec:merging}).}
  \label{tab:timingsA}
  \bigskip
  \begin{tabular}{l*{5}{|r}}
    \hline
     & $d = -4$ & $d = -3$ & $d = -2$ & $d = -1$ & $d = 0$ \\ 
Panjer & 0.461 & 40.784 & 4651.298 &   NA &   NA \\ 
MonteCarlo & 1.246 & 2.085 & 5.820 & 10.228 & 12.413 \\ 
Moment & 0.011 & 0.006 & 0.010 & 0.019 & 0.025 \\ 
Chernoff & 0.112 & 0.310 & 0.634 & 1.017 & 1.284 \\ 
Cantelli & 0.001 & 0.002 & 0.001 & 0.002 & 0.002 \\ 
Markov & 0.001 & 0.001 & 0.001 & 0.005 & 0.001 \\ 

    \hline\hline
  \end{tabular}
\end{table}

A similar table to Table~\ref{tab:timingsA} could be constructed for
any specified value $s_0$, rather than a whole set of values.  The
timings for the Moment, Chernoff, Cantelli, and Markov bounds would
all be roughly one hundredth as large, because these are evaluated
pointwise.  The timing for Monte Carlo would be unchanged.  The timing
for Panjer would be roughly the proportion $s_0 / \$40 \unit{m}$ of
the total timing, because it is evaluated sequentially, from small to
large values of~$s$.

\subsection{Gamma thickening of the event losses}
\label{sec:thickening}

\begin{figure}
  \centering
  \includegraphics[width=\textwidth]{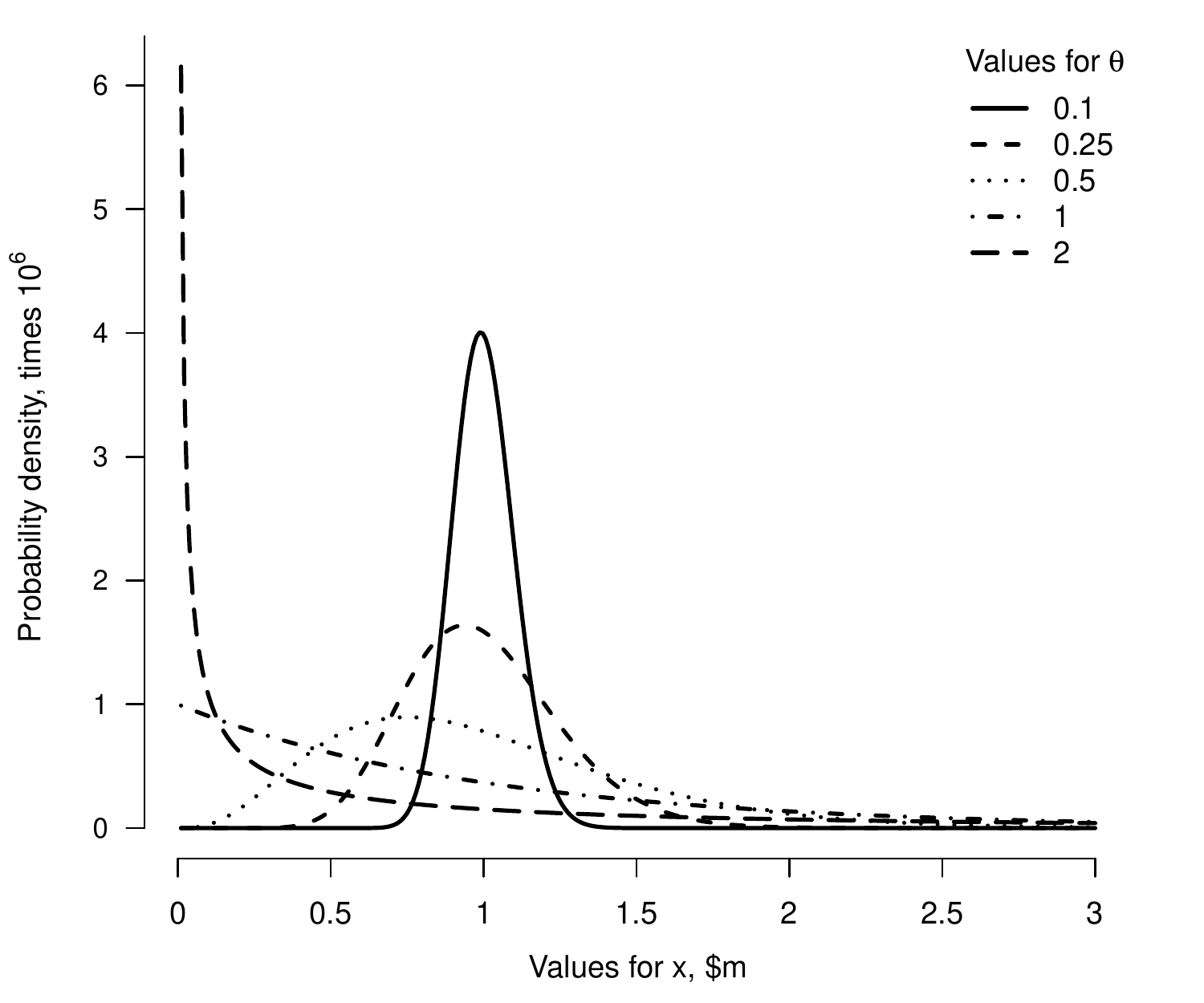}
  \caption{Effect of varying $\theta$ on the shape of the Gamma
    distribution with expectation $\$1 \unit{m}$.}
  \label{fig:thetas}
\end{figure}

We continue to consider one-year losses, but now treat the losses from
each event as random, not fixed.  For the simplest possible
generalisation we use a Gamma distribution with a specified
expectation $x_i$ and a common specified coefficient of variation,
$\theta \ldef \sigma_i / x_i$.  The previous case of a fixed loss
$x_i$ is represented by $\lim \theta \to 0$, which we write,
informally, as $\theta = 0$.  Solving
\begin{displaymath}
  x_i = \frac{\alpha_i}{\beta_i} \quad \text{and} \quad \theta x_i = \sqrt{ \frac{\alpha_i}{\beta_i^2} }
\end{displaymath}
gives the two Gamma distribution parameters as
\begin{displaymath}
  \alpha_i = \frac{1}{\theta^2} \quad \text{and} \quad \beta_i = \frac{\alpha_i}{x_i} .
\end{displaymath}
Figure~\ref{fig:thetas} shows the effect of varying $\theta$ on a
Gamma distribution with expectation $\$1 \unit{m}$.

The only practical difficulty with allowing random losses for each
event occurs for the Panjer method; we describe our approach in the
Appendix.

\begin{figure}[p]
  \includegraphics[width=\textwidth]{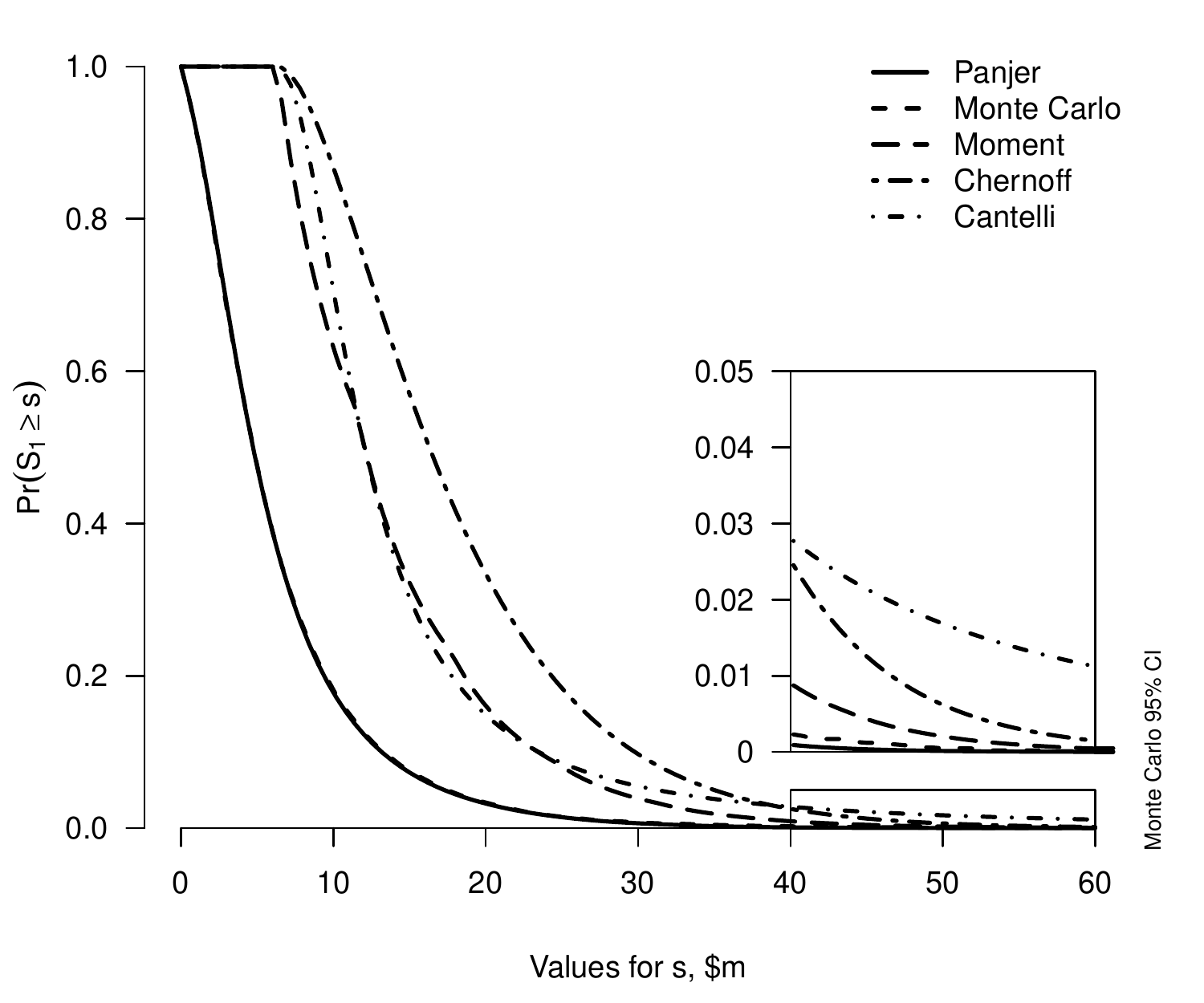}
  \caption{As Figure~\ref{fig:EPA}, with $t = 1$, $u = \infty$, $\theta
    = 0.5$, and $d = -4$.  The Markov bound has been dropped.  Timings
    are given in Table~\ref{tab:timingsB}.}
  \label{fig:EPB}
\end{figure}

\begin{table}
  \centering\renewcommand{\arraystretch}{1.25}
  \caption{Timings for the  methods shown in Figure~\ref{fig:EPB}.}
  \label{tab:timingsB}
  \bigskip
  \begin{tabular}{l*{5}{|r}}
    \hline
     & $d = -4$ & $d = -3$ & $d = -2$ & $d = -1$ & $d = 0$ \\ 
Panjer & 1.509 & 121.062 &   NA &   NA &   NA \\ 
MonteCarlo & 0.921 & 1.961 & 5.385 & 9.967 & 12.197 \\ 
Moment & 0.006 & 0.021 & 0.055 & 0.100 & 0.118 \\ 
Chernoff & 0.127 & 0.614 & 1.670 & 2.734 & 3.333 \\ 
Cantelli & 0.001 & 0.002 & 0.007 & 0.022 & 0.019 \\ 

    \hline\hline
  \end{tabular}
\end{table}

Figure~\ref{fig:EPB} shows the exceedance probability curve with
$\theta = 0.5$: note that the horizontal scale now covers a much wider
range of loss values than Figure~\ref{fig:EPA}.  The timings are given
in Table~\ref{tab:timingsB}: these are very similar to the non-random
case with $\theta = 0$ (Table~\ref{tab:timingsA}), with the exception
of the Panjer method, which takes longer because it scales linearly
with the upper limit on the horizontal axis.

\subsection{Capping the loss from a single event}
\label{sec:capping}

Now consider the case where the single-event loss is capped at $\$5
\unit{m}$.  The implementation of this cap is straightforward, and we
describe it in the Appendix.  The results are given in
Figure~\ref{fig:EPC} and Table~\ref{tab:timingsC}.  For the timings,
the main effect of the cap is on the Panjer method, because the cap
reduces the probability in the righthand tail of the loss
distribution, and allows us to use a smaller upper limit on the
horizontal axis.  But the Panjer approximation, where it can be
computed, still takes a thousand times longer to compute than the
Moment bound.

\begin{figure}[p]
  \includegraphics[width=\textwidth]{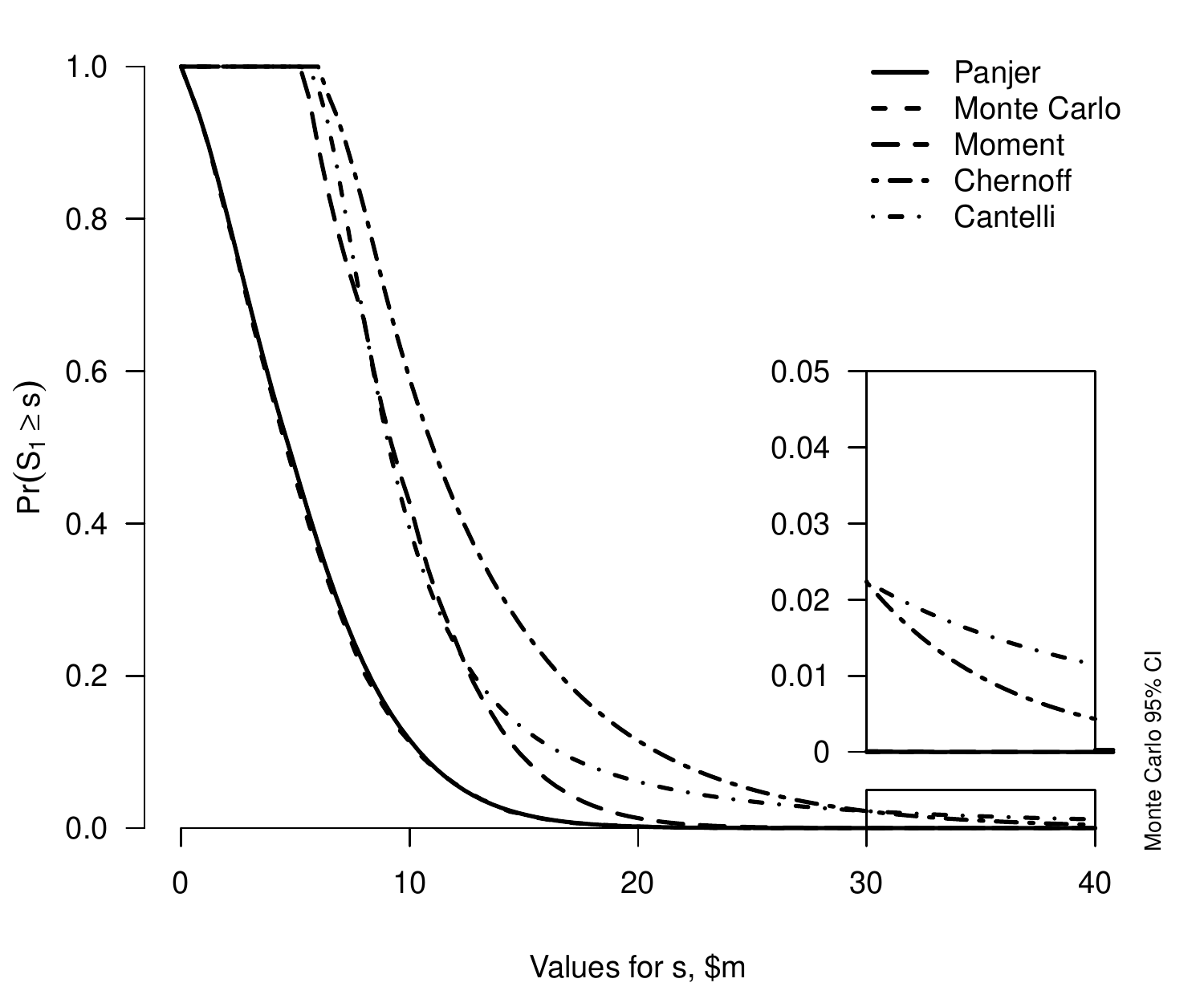}
  \caption{As Figure~\ref{fig:EPA}, with $t = 1$, $u = \$5 \unit{m}$, $\theta
    = 0.5$, and $d = -4$. Timings
    are given in Table~\ref{tab:timingsC}.}
  \label{fig:EPC}
\end{figure}

\begin{table}
  \centering\renewcommand{\arraystretch}{1.25}
  \caption{Timings for the  methods shown in Figure~\ref{fig:EPC}.}
  \label{tab:timingsC}
  \bigskip
  \begin{tabular}{l*{5}{|r}}
    \hline
     & $d = -4$ & $d = -3$ & $d = -2$ & $d = -1$ & $d = 0$ \\ 
Panjer & 0.275 & 11.950 &   NA &   NA &   NA \\ 
MonteCarlo & 1.106 & 2.099 & 5.625 & 10.329 & 11.837 \\ 
Moment & 0.016 & 0.070 & 0.210 & 0.355 & 0.431 \\ 
Chernoff & 0.508 & 2.057 & 5.659 & 10.853 & 14.992 \\ 
Cantelli & 0.003 & 0.006 & 0.019 & 0.032 & 0.041 \\ 

    \hline\hline
  \end{tabular}
\end{table}

\subsection{Ten-year losses}
\label{sec:tenyears}

\begin{figure}[p]
  \includegraphics[width=\textwidth]{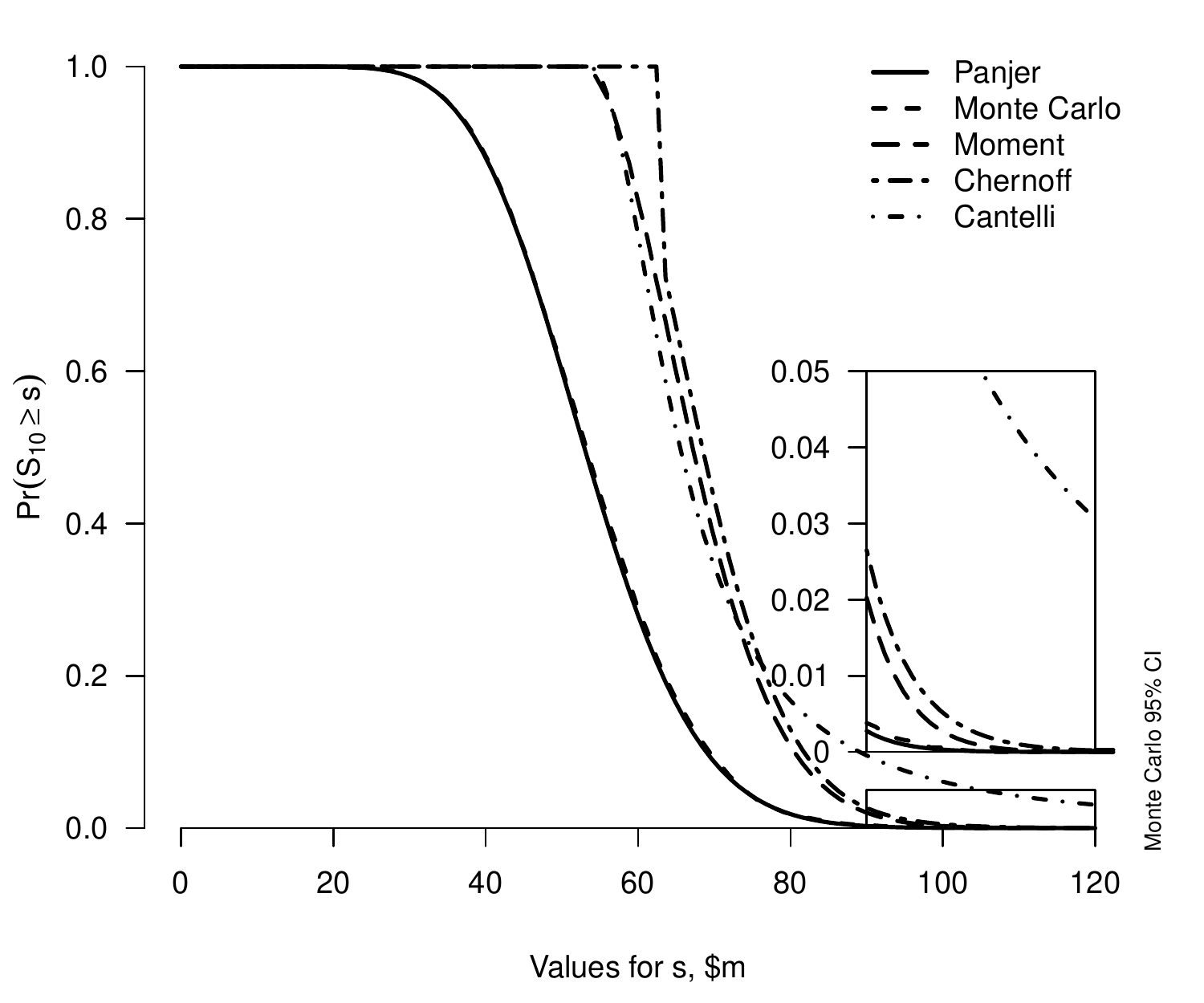}
  \caption{As Figure~\ref{fig:EPA}, with $t = 10$, $u = \$5 \unit{m}$, $\theta
    = 0.5$, and $d = -4$. Timings
    are given in Table~\ref{tab:timingsD}.}
  \label{fig:EPD}
\end{figure}

Finally, consider expanding the time period from $t = 1$ to $t =
10$~years; the results are given in Figure~\ref{fig:EPD} and
Table~\ref{tab:timingsD}.  The timings of the Markov, Cantelli,
Moment, and Chernoff bounds are unaffected by the value of $t$.  The
timing for the Panjer method grows with $t$, because the righthand
tail of $S_t$ grows with $t$. The timing for the Monte Carlo method
grows roughly linearly with $t$, but the `in simulation' time for
Monte Carlo is dominated by other factors, so the additional computing
time for the increase in $t$ from $t = 1$ to $t = 10$, is small.

\begin{table}
  \centering\renewcommand{\arraystretch}{1.25}
  \caption{Timings for the  methods shown in Figure~\ref{fig:EPD}.}
  \label{tab:timingsD}
  \bigskip
  \begin{tabular}{l*{5}{|r}}
    \hline
     & $d = -4$ & $d = -3$ & $d = -2$ & $d = -1$ & $d = 0$ \\ 
Panjer & 0.587 & 46.000 &   NA &   NA &   NA \\ 
MonteCarlo & 918.101 & 2.312 & 7.189 & 11.834 & 14.869 \\ 
Moment & 0.027 & 0.141 & 0.435 & 0.592 & 0.736 \\ 
Chernoff & 3.989 & 2.101 & 7.794 & 10.503 & 18.314 \\ 
Cantelli & 0.002 & 0.007 & 0.020 & 0.164 & 0.046 \\ 

    \hline\hline
  \end{tabular}
\end{table}

\section{\MakeUppercase{Summary}}
\label{sec:summary}

We have presented four upper bounds and two approximations for the
upper tail of the loss distribution that follows from an Event Loss
Table (ELT).  We argue that in many situations an upper bound on this
probability is sufficient.  For example, to satisfy the regulator, in
a sensitivity analysis, or when there is supporting evidence that the
bound is quite tight.  Of the bounds we have considered, we find that
the Moment bound offers the best blend of tightness and computational
efficiency.  In fact, the Moment bound is effectively costless to
compute, based on the timings from our \textsf{R} package.

We have stressed that there are no exact methods for computing tail
probabilities when taking into account limited computing resources.
Of the approximately exact methods we consider, we prefer Monte Carlo
simulation over Panjer recursion, because of the availability of an
error estimate in the former and the amount of information provided by the latter.
A back-of-the-envelope calculation
suggests that 10,000 Monte Carlo simulations should suffice to satisfy
the Solvency~II regulator.

The merging operation is a very useful way to condense an ELT that has become bloated, for example after using mixtures of Gamma distributions to represent more complicated secondary uncertainty distributions.  We have shown that the Moment bound provides a quick way to assess how much merging can be done without having a major impact on the resulting aggregate loss distribution.

We have also demonstrated the versatility of the Gamma distribution
for single event losses.  The Gamma distribution has a simple moment
generating function and explicit expressions for the moments.
Therefore it fits very smoothly into the compound Poisson process that
is represented in an ELT, for the purposes of computing approximations
and bounds.  We also show how the Gamma distribution can easily be
adapted to account for a cap on single event losses.  We favour the
capped Gamma distribution over the Beta distribution, which is often
used in the industry, because the former has an atom (as is
appropriate) while the latter does not.

\section*{\MakeUppercase{Acknowledgements}}

We would like to thank Dickie Whittaker, David Stephenson and Peter Taylor for valuable comments which helped in improving the exposition of this paper, and for supplying the US Hurricanes event loss table.  This work was funded in part by NERC grant~NE/J017450/1, as part of the CREDIBLE consortium.

\section*{\MakeUppercase{Appendix}}

\paragraph{Minimisation for the Moment bound.}

The Moment bound is minimised over the control variable $k = 1, 2,
\dots$.  It is convenient to have an upper bound for $k$, because it
is efficient to compute $S_t^k$ for a set of $k$ values, rather than one
$k$ at a time, as shown in \eqref{eq:cpp}.  We find an approximate
upper bound for $k$ as follows.  First, we compute the first two
moments of $S_t$ exactly using \eqref{eq:Smom1} and \eqref{eq:Smom2}.
Then we approximate the distribution of $S_t$ using a Gamma
distribution matched to these two moments, for which
\begin{displaymath}
  \alpha_s =  \frac{ \mu_t^2 }{ \sigma_t^2 } \qquad \beta_s = \frac{ \mu_t }{ \sigma_t^2 } .
\end{displaymath}
The moments for the Gamma distribution were given in
\eqref{eq:gammamom}.  Starting from this expression and $k = 1$, we
step out in $k$ until the Gamma approximation to $\log \E(S_t^k) / s^k$
shows an increase on its previous value.  The ceiling of the resulting
$k$ is taken as the maximum $k$ value.  If the Moment bound is
required for a sequence of $s$ values, we use the largest $s$ value in
the sequence.

\paragraph{Minimisation for the Chernoff bound.}

The MGF for a Gamma distribution is given in \eqref{eq:MGFgamma}.
Hence the range for the control variable $v$ is $0 < v < \min_i \{
\beta_i \}$.  As explained in section~\ref{sec:thickening}, we specify
the two parameters of the Gamma distribution for event $i$ in terms of
the fixed loss $x_i$, now treated as the expected loss, and a
coefficient of variation $\theta$ (which could vary with $i$).  This
gives $\beta_i = 1 / (\theta^2 x_i)$, and hence
\begin{displaymath}
  v < \min_i \left\{ \frac{1}{ \theta^2 x_i} \right\} .
\end{displaymath}
In the simpler case of a fixed loss for event $i$, we substitute the
small coefficient of variation, $\theta = 0.1$, to give $v < \min_i \{
100 / x_i \}$.  We perform the minimisation over a set of $1001$
equally-spaced values for $v$.

\paragraph{Panjer recursion for random event losses.}

The Panjer algorithm needs each event loss to be a fixed
(non-negative) integer.  Therefore we follow the mixture approach of
section~\ref{sec:special} to replace an event $i$ with a random loss with
a collection of events with fixed losses.  Consider event $i$, with
loss distribution $f_i$.  We replace row $i$ in the original ELT with
$n_q$ rows each with rate $\lambda_i / n_q$, and with losses
$v^{(i)}_1, \dots, v^{(i)}_{n_q}$, where $v^{(i)}_j$ is the
\begin{displaymath}
  \left( \frac{j}{n_q} - \frac{1}{2 n_q} \right) \text{th}
\end{displaymath}
quantile of $f_i$.  Having done this for all rows, we then compress
the expanded ELT back to integer values again (i.e.\ using ${d = 0}$).
We used $n_q = 10$.

\paragraph{Capping single event losses.}

In the case where event losses are non-random, a cap at $u$ simply
replaces each loss $x_i$ for which $x_i > u$ with the value $u$.
Where the event losses are Gamma-distributed with expectation $x_i$
and specified coefficient of variation $\theta$, the modified Gamma
moment generating functions are used for the Markov, Cantelli, Moment,
and Chernoff method, see section~\ref{sec:special}.  The Panjer method
is implemented on an augmented ELT, as described immediately above,
and then each loss is capped at $u$.  The Monte Carlo method has each
sampled loss capped at $u$.


\begin{thebibliography}{10}
\expandafter\ifx\csname natexlab\endcsname\relax\def\natexlab#1{#1}\fi

\bibitem[Brown{\em \ et~al.}(2001)Brown, Cai, and DasGupta]{brown01}
L.D. Brown, T.T. Cai, and A.~DasGupta, 2001.
\newblock Interval estimation for a binomial proportion.
\newblock {\em Statistical Science}, {\bf 16}\penalty0 (2), \penalty0 101--117.
\newblock With discussion, pp~117--133.

\bibitem[Grimmett and Stirzaker(2001)]{grimmett01}
G.R. Grimmett and D.R. Stirzaker, 2001.
\newblock {\em Probability and Random Processes}.
\newblock Oxford, UK: Oxford University Press, 3rd edition.

\bibitem[Hickey and Hart(2013)]{hickey13}
G.L. Hickey and A.~Hart, 2013.
\newblock Statistical aspects of risk characterisation in ecotoxicology.
\newblock In J.C. Rougier, R.S.J. Sparks, and L.J. Hill, editors, {\em Risk and
  Uncertainty Assessment for Natural Hazards}, chapter~14. Cambridge University
  Press, Cambridge, UK.

\bibitem[Philips and Nelson(1995)]{philips95}
T.K. Philips and R.~Nelson, 1995.
\newblock The moment bound is tighter than {Chernoff's} bound for positive tail
  probabilities.
\newblock {\em The American Statistician}, {\bf 49}\penalty0 (2), \penalty0
  175--178.

\bibitem[{R Core Team}(2013)]{Rproject}
{R Core Team}.
\newblock {\em R: A Language and Environment for Statistical Computing}.
\newblock R Foundation for Statistical Computing, Vienna, Austria, 2013.

\bibitem[Ross(1996)]{ross96}
S.M. Ross, 1996.
\newblock {\em Stochastic Processes}.
\newblock John Wiley \& Sons, Inc., New York, USA, second edition.

\bibitem[Shevchenko(2010)]{shev10}
P.V. Shevchenko, 2010.
\newblock Calculation of aggregate loss distributions.
\newblock {\em The Journal of Operational Risk}, {\bf 5}\penalty0 (2),
  \penalty0 3--40.

\bibitem[Whittle(2000)]{whittle00}
P.~Whittle, 2000.
\newblock {\em Probability via Expectation}.
\newblock New York: Springer, 4th edition.

\bibitem[Wilkinson(2012)]{wilkinson12}
D.J. Wilkinson, 2012.
\newblock {\em Stochastic Modelling for Systems Biology}.
\newblock CRC Press, Boca Raton FL, USA, second edition.

\bibitem[Wiper{\em \ et~al.}(2001)Wiper, Insua, and Ruggeri]{Wiper01}
Michael Wiper, David~Rios Insua, and Fabrizio Ruggeri, 2001.
\newblock Mixtures of {Gamma} distributions with applications.
\newblock {\em Journal of Computational and Graphical Statistics}, {\bf
  10}\penalty0 (3), \penalty0 440--454.

\end{thebibliography}

\end{document}